\def\ref{\par\noindent\hang}
\def\AaA{A\&A}

\def\ApJ{{ApJ}}
\def\ApJL{{ApJ}}

\def\ARAA{ARA\&A}

\def\Nat{{Nat}}

\def\etal{{\it et al.\ }}

\def\km{{\rm\thinspace km}}

\def\s{{\rm\thinspace s}}

\def\kmps{\hbox{$\km\s^{-1}\,$}}

\def\1E{1E~1415.6+2557}

\documentstyle[aas2pp4]{article}

\begin{document} \title{The Distance to the Draco Cloud}

\author{ Michael D. Gladders, T. E. Clarke,
  Christopher R. Burns, A. Attard, M. P. Casey, Devon
  Hamilton, Gabriela Mall\'{e}n-Ornelas, J. L. Karr, Sara M. Poirier,
  Marcin Sawicki, L. Felipe Barrientos and Stefan W. Mochnacki}
\affil{Department of Astronomy and The David Dunlap Observatory,
  60 St. George Street, University of Toronto, M5S 3H8, Canada}
\authoremail{gladders@astro.utoronto.ca}

\begin{abstract} \noindent The understanding of the nature of intermediate 
  and high velocity gas in the Milky Way is hampered by a paucity of
  distance estimates to individual clouds. A project has been started
  at the David Dunlap Observatory to address this lack of distance
  measures by observing early-type stars along the line of sight
  towards these clouds and searching for sodium doublet absorption at
  the clouds' systemic velocities. Distances to foreground stars (no
  absorption) and background stars (with absorption) are estimated
  from spectroscopic parallax, and thus the distance to the bracketed
  cloud is estimated.  In this Letter, we present the first result
  from this ongoing project, a measurement of the distance to the
  Draco Cloud, which is the most studied of the intermediate velocity
  clouds. The result presented here is the first distance bracket
  which tightly constrains the position of the Draco Cloud.  We
  briefly describe our target selection and observing methodology, and
  then demonstrate absorption at the velocity of the Draco Cloud for
  one star (TYC~4194~2188), and a lack of absorption for several other
  stars.  We derive a distance bracket to the Draco Cloud of
  463$^{+192}_{-136}$ to 618$^{+243}_{-174}$~pc.
\end{abstract}


\section{INTRODUCTION}
\noindent Observations in the 21cm HI line reveal the presence of discrete clouds
of neutral hydrogen at high galactic latitudes (e.g. Hartmann \&
Burton 1997). These clouds, whose velocities cannot be explained by
differential galactic rotation, are generally grouped into velocity
subsets based on their radial velocities relative to the local
standard of rest (LSR). Historically, the clouds
with $|V_{LSR}|$ $>$ 70 \kmps are labeled high-velocity clouds (HVCs)
and those with $|V_{LSR}|$ $<$ 70 \kmps are termed
intermediate-velocity clouds (IVCs). Recent indications from both dust
surveys and distance estimates suggest that the two classes are indeed
distinct populations (Burton 1996).

The distances to the clouds are necessary to constrain the most
significant physical parameters, such as mass, size, density and
pressure. Without distance information, the origins of these clouds
remains speculative.  Numerous origin scenarios have been proposed,
including Local-Group infall from the intergalactic medium (Blitz
\etal 1998), products of thermal instabilities acting at large scale
heights in the galactic halo, and cooled infalling gas which was
ejected into the halo from energetic disk events (see review by Wakker
\& van Woerden 1997). It is essential to determine the cloud distances
in order to distinguish between such scenarios.

Cloud distances are also fundamental to understanding the soft X-ray
background (SXRB). For example, SXRB measurements in the direction of
the Draco Cloud show a reduction of $\sim$ 50\% in the 0.25~keV
emission over the cloud relative to the adjacent sky (Burrows \&
Mendenhall 1991; Snowden \etal 1991). This depression is consistent
with photoelectric absorption of the SXRB emission by the cloud. The
nature of the SXRB is uncertain. One model suggests that most of the
SXRB arises from local low-density plasma (Saunders \etal 1977). This
Local Hot Bubble (LHB) has a temperature of $\sim 10^6$~K and extends
for a few hundred parsecs. Other models propose that the majority of
the SXRB emission originates at large distances in the Galactic halo
(e.g. Marshall and Clark 1984). Again, in order to distinguish between
models, it is essential to determine the distances to the shadowing
clouds.

The Draco Cloud ($\ell$=90$^{\circ}$.0, b=38$^{\circ}$.8) is the best
studied of the high-latitude IVCs (e.g. Lilienthal \etal 1991 and
references therein), but its distance is still uncertain.  Two
different upper distance limits have been suggested by: 1) star count
measurements (Rohlfs 1986: 2000~pc) and 2) differential star count
measurements (Wennmacher 1988: 1500~pc). Three different lower
distance limits have been suggested by: 1) $UBV$ photometry (Goerigk \&
Mebold 1986: 800~pc), 2) star count measurements (Mebold \etal 1985:
800~pc) and 3) interstellar absorption-line measurements (Lilienthal
\etal 1991: 300~pc). Notably, the limits from star count data and
$UBV$ photometry may require significant revision, as recent work by
Stark \etal (1997) indicates that the extinction due to the Draco
Cloud may have been overestimated by a factor of four or more. Of all
methods used, the most promising is interstellar absorption line
measurements, both in terms of the potential accuracy of the derived
distance, and the insensitivity to systematic errors (Wakker \& van
Woerden 1997).

In this Letter, we derive a distance to the Draco Cloud using the
interstellar absorption-line technique. This distance estimate is
based on moderate-resolution spectra at the sodium doublet of four stars
in the line of sight to this cloud. One star clearly displays
interstellar absorption at the systemic velocity of the Draco Cloud
and is thus behind it. The three other stars show no interstellar
absorption, and are thus in front of it. From spectral
classifications, we deduce a distance to each star, and hence bracket
the cloud distance. This represents the first measurement of the
distance to the Draco Cloud where both firm upper and lower limits can
be given. In \S2 we describe the target selection process,
observations and data reduction. In \S3 we demonstrate the
interstellar absorption in the background star, and the lack of
interstellar absorption in the foreground stars. The distance to the
Draco Cloud is derived in \S4.

\section{OBSERVATIONS}
\noindent Numerous stars along the line-of-sight towards the Draco
Cloud were observed at the sodium doublet (5889.953 \AA~\& 5895.923
\AA) from May to September, 1997, and in February 1998, using the David Dunlap
Observatory (DDO) 1.88m telescope + Cassegrain spectrograph. 
The setup used provided a dispersion of 0.200
\AA/pixel, a spectral range of $\sim 200$ \AA, with a resolution of 0.43 \AA. This resolution, equivalent
to 22 \kmps at the wavelength of the sodium doublet, is
sufficient to detect the sharp lines attributable to individual HI
clouds (Wakker \& van Woerden 1997). 

The target stars were selected primarily from the Tycho catalog (Hog
\etal 1997), the deepest available general catalog of stars with
accurate $B-V$ colors. We selected stars which are blue ($B-V<0.4$),
and relatively faint ($V>9.0$). Such stars are likely to be distant
early-type stars, which provide relatively clean continuum emission in
the spectral region of interest, and are thus useful as probes for
interstellar absorption. The bulk of the observing time was allocated
to stars which clearly lay along the line of sight to the Draco Cloud
as defined by it's IRAS $dust$ emission at 100$\mu$m (Figure 1). This
dust should trace the highest column density regions of the cloud
(Burton 1996).  We expect such stars to show a relatively strong
sodium absorption feature (if behind the cloud), easily detected as
unresolved lines in moderate resolution spectra. This selection
technique is in contrast to previous work on other IVCs ({\it e.g.}
Benjamin \etal 1996), where targets were defined on the basis of the
more extended HI emission, and showed a correspondingly weaker sodium
absorption.
    
A total of eight stars were observed toward the Draco Cloud, with
typical total integration times ranging from 2400-5400 seconds.
Telluric standards were repeatedly observed at regular intervals. Of
the six standards observed, only HD177724 and HD120315 were eventually
used, as the other standards showed contamination by interstellar
sodium absorption at low velocities. The target star observations are
detailed in Table 1.  One star (BD+64 1134) was extensively
re-observed in February 1998, as the original 1997 spectra showed a
potential, albeit weak, sodium absorption feature at the Draco Cloud's
systemic velocity; this feature was not present in the final co-added
spectrum.

The spectra were pre-processed and extracted to \mbox{1-D} using
standard IRAF\footnote{IRAF is distributed by the National Optical
  Astronomy Observatories, which is operated by the Association of
  Universities for Research in Astronomy, Inc., under contract to the
  National Science Foundation.} tasks.  Great care was taken to ensure
the correct removal of prominent sodium sky lines engendered by the
DDO's close proximity to a major urban center. The individual spectra
for each target star were then summed after velocity correction to the
LSR rest frame. In the following analysis, we concentrate on
the four most distant stars (based on color); we also list 
the four closer stars in Table 1 for completeness.

\section{FOREGROUND AND BACKGROUND STARS}

The spectra for the four most distant stars are shown in Figure 2.
One of these stars (TYC~4194~2188) $clearly$ shows an {\it unresolved}
absorption feature at the Draco Cloud's systemic velocity
($V_{LSR}$=--23.9 \kmps~; Heiles, Reach \& Koo 1988), and is thus
behind the cloud. To quantify the nature of the other strong
sodium absorption features apparent in the spectra, we have cross-correlated
the {\it remaining} wavelength portions of the spectra with an appropriate
spectral standard (HD173667). The cross-correlations show maxima
which have velocities and widths consistent with all other significant
sodium absorption features, indicating that these features are likely
stellar in origin. The LSR velocities of these stellar features are
listed in Table 2.  Furthermore, we have used the observed
interstellar absorption in TYC~4194~2188 (equivalent width $\approx$
0.25~\AA) to compute the expected interstellar absorption in the other
three stars, presuming that these stars are also behind the Draco
Cloud.  This putative absorption has been scaled by the relative total
HI column density (due to the Draco Cloud) at each star compared to
that at TYC~4194~2188, derived from the Leiden/Dwingeloo HI survey
(Hartmann \& Burton 1997). The expected absorption is shown in Figure
2. Since their observed spectra do not show the expected absorption,
we claim that the other three stars are in front of the Draco Cloud.
Note that this assumes that the NaI$/$HI ratio remains constant across
the cloud, and that changes in the HI column density at angular scales
less than $\sim$0.5 degrees do not significantly affect our
conclusions.

\section{DISTANCE DETERMINATIONS}
To determine the distance to this subset of four stars, high S/N
`classification' spectra were acquired, again using the DDO~1.88m
telescope + Cassegrain spectrograph. The spectra cover roughly
3750-4350~\AA. A suite of standards (from Garcia 1989) were also
observed, ranging in spectral type from B0 to G5, and spanning luminosity
classes I-V.

The four stars with classification spectra have been classified by
comparison to the observed set of spectral standards.  The derived
spectral class and the absolute magnitude calibration of Corbally \&
\mbox{Garrison} (1984) were then used to assign an absolute magnitude,
and hence distance, to each star (see Table 2). We have considered
three independent sources of error in this distance estimate: 1) the
error in the Tycho $V$ magnitude; 2) an assumed classification error
of $\pm$ one spectral class for each star; 3) the intrinsic dispersion
(0.7~mag) in the absolute magnitude relation vs. MK spectral class
(Jaschek \& G\'omez 1998). The dominant uncertainty is the intrinsic
scatter in the M$_{V}$-MK relation. We have not accounted for
extinction, as the expected overall extinction along typical lines of
sight at these galactic latitudes is small. Additionally, the
extinction on the background star due to the Draco Cloud is uncertain,
but likely quite small (Stark \etal 1997). Note that the inclusion of
extinction can only {\it reduce} the estimated distance of the
background star, and hence tighten the distance bracket.

\begin{center}{\it Individual Stars}\end{center}
\noindent{\it TYC~4194~2188:} This star, the only one of our sample with
detected interstellar sodium absorption due to the Draco Cloud, is
also obviously of the earliest spectral type. The best spectral class
estimate is A6V, which places it at a distance of $\sim$620~pc.\nl
\noindent{\it BD+64~1134:} This star is an F5 main-sequence dwarf. Though it is
quite distant ($\sim$380~pc), the weak expected sodium absorption (Figure 2)
makes it a poor lower limit, despite the relatively high S/N of
the sodium spectrum.\nl
\noindent{\it TYC~4190~1263:} This star is a spectroscopic binary as
it shows double stellar lines in the sodium spectral region. The flux
ratio of the components is $\sim$2-3, and the best composite spectral
classification is an F5/G0 main-sequence dwarf pair. In determining
the distance to this star, we have allowed a spectral mis-classification of
$\pm$1 class for the primary component and $\pm$2 classes for the
secondary. This star provides the best lower limit on the Draco Cloud
as it is relatively distant ($\sim$460~pc), and would be expected to
readily show sodium absorption if it were behind the cloud (see Figure 2).\nl
\noindent{\it TYC~4193~691:} This star is about G0, and likely a
main-sequence dwarf. The angular separation between it and the background
star is quite small, but it is too nearby ($\sim$250~pc) to provide a useful
lower distance limit.\nl

The derived distances are summarized in Table~2.  The background star
(TYC~4194~2188) and the most distant foreground star (TYC~4190~1263)
set constraints on the distance to the Draco Cloud. This bracket is
463$^{+192}_{-136}$ to 618$^{+243}_{-174}$~pc, which corresponds to a
vertical distance above the galactic plane of $290^{+120}_{-85}$ to
$387^{+152}_{-109}$~pc. This is in excellent overall agreement with
the much broader distance limits set by other methods (see \S1).

\section{CONCLUSIONS}
We have observed a total of 8 stars in the line of sight towards the
Draco Cloud and detected the cloud in absorption at the sodium doublet
toward one star. Detailed spectral classifications of this star and the
three most distant foreground stars provide a distance bracket of
463$^{+192}_{-136}$ to 618$^{+243}_{-174}$~pc. This distance is
consistent with, but much more precise than, other previous measures
by different means. Furthermore, this distance indicates that the
Draco Cloud is located outside the Local Hot Bubble and therefore must
be shadowing X-ray emission from beyond the LHB.

\centerline{ACKNOWLEDGMENTS}
M.D.G, C.R.B. and A.A. wish to thank the Natural Sciences and Engineering Research
Council of Canada for support through the PGS~A and PGS~B graduate
scholarship programs. We also wish to acknowledge the immense support
this project has received from the Director and staff of the David
Dunlap Observatory. We thank, in particular, J. Thomson and W. Lu, the
two telescope operators who have vigorously supported our research
program. Additionally, we would like to acknowledge the contributions
of the late Karl Kamper, whose enthusiasm and support were of great
value to us. Finally, we are grateful to both Bob Benjamin and Bob
Garrison for many useful discussions regarding IVCs and MK
classification, respectively.

\vfill
\eject

\centerline{REFERENCES}

\ref{Benjamin, R. A., Venn, K. A., Hiltgen, D. D., \& Sneden, C. 1996, \ApJ,
  464, 836}
\ref{Blitz, L., Spergel, D. N., Teuben, P. J., Hartmann, D., \& Burton,
  W. B. 1998, {\it preprint} (astro-ph/9803251)}
\ref{Burrows, D. N. \& Mendenhall, J. A. 1991, \Nat, 351, 629}
\ref{Burton, W. B. 1996, in {\it The Physics of Galactic Halos}, eds. H. Lesch,
  , R.-J. Dettmar, U. Mebold, and R. Schlickeiser, (Berlin: Akademie Verlag), p. 15}
\ref{Corbally, C. J., \& Garrison, R. F. 1984, in {\it The MK Process and
    Stellar Classification}, ed. R. F. Garrison, (Toronto: David Dunlap
  Observatory), p. 277}
\ref{Garcia, B. 1989, BICDS, 36, 27}
\ref{Goerigk, W., \& Mebold, U. 1986, \AaA, 162, 279}
\ref{Hartmann, D., \& Burton, W. B. 1997, {\it Atlas of Galactic Neutral
  Hydrogen}, (Cambridge: Cambridge Univ. Press)}
\ref{Heiles, C., Reach, W. T., \& Koo, B. 1988, \ApJ, 332, 313}
\ref {Hog, E., Baessgen G., Bastian, U., Egret, D., Fabricius, C.,
  Grossman, V., Halbwachs, J. L., Makarov, V. V., Perryman, M. A. C,
  Schwekendiek, P., Wagner, K., \& Wicenec, A. 1997, \AaA, 323, 57}
\ref{Jaschek, C., \& G\'omez, A. E. 1998, \AaA, 330, 619}
\ref{Lilienthal, D., Wennmacher, A., Herbstmeier, U., \& Mebold, U.
  1991, \AaA, 250, 150}
\ref{Marshall, F. J., \& Clark, G. W. 1984, \ApJ, 287, 633}
\ref{Mebold, U., Cernicharo, J., Velden, L., Reif, K., Crezelius, C., \&
  Goerigk, W. 1985, \AaA, 151, 427}
\ref{Rohlfs, R. 1986, Diploma Thesis, University of Bonn}
\ref{Saunders, W. T., Kraushaar, W. L., Nousek, J. A., \& Fried, P. M. 1977, \ApJL, 217, L87}
\ref{Snowden, S. L., Mebold, U., Hirth, W., Herbstmeier, U., \& Schmitt, J. H. M. M. 1991, Science, 252, 1529}
\ref{Stark, R., Kalberla, P., \& G\"usten, R. 1997, \AaA, 317, 907}
\ref{Wakker, B. P., \& van Woerden, H. 1997, \ARAA, 53, 217}
\ref{Wennmacher, A. 1988, Diploma Thesis, University of Bonn}
\ref{Wheelock, S. \etal 1994, {\it IRAS Sky Survey Atlas Explanatory Supplement}, JPL Publication 94-11, (Pasadena: JPL)}

\vfill
\eject

\begin{figure}[h]
\plotone{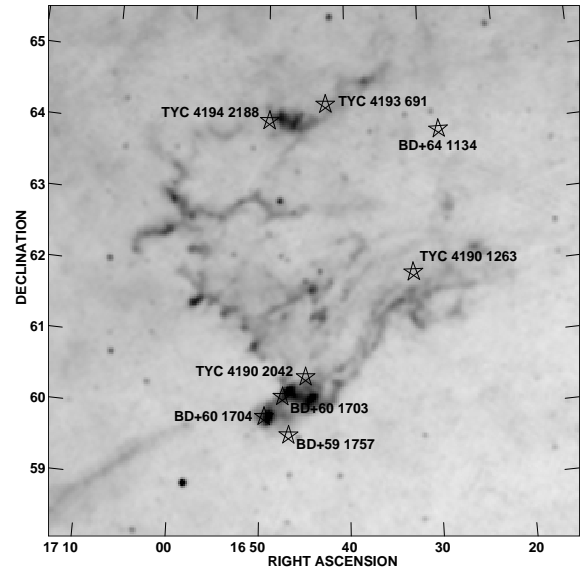}
\caption{The IRAS 100$\mu$m image of the Draco Cloud (Wheelock \etal 1994)
  with the positions of the 8 program stars shown.}
\end{figure}
 
\begin{figure}[h]
\plotone{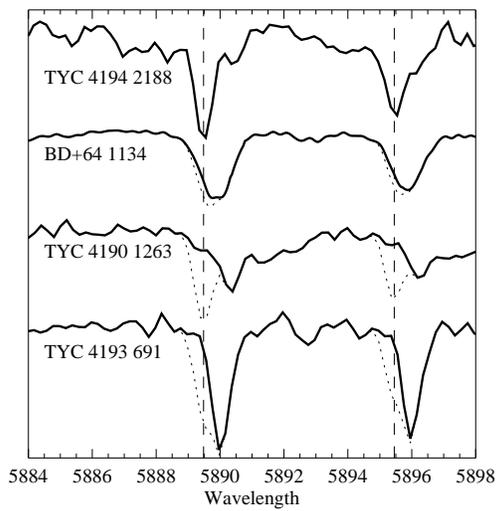}
\caption{Moderately high resolution spectra (in arbitrary linear intensity units), corrected to the LSR frame, for the four
most distant program stars (thick lines) showing both stellar and interstellar
sodium absorption features. Note the
prominent absorption feature in the spectrum of TYC~4194~2188, at
the velocity of the Draco Cloud (dashed vertical lines). The thin
dotted lines show the expected effect of similar absorption on the other
three stars, as described in \S3.}
\end{figure}

\vfill
\eject

\begin{deluxetable}{cccc}
\tablewidth{0pc}
\tablecaption{Stars Observed at the Sodium Doublet}
\tablehead{
\colhead{Star}&
\colhead{V}&
\colhead{B-V}&
\colhead{Int.}\\
\colhead{}&
\colhead{}&
\colhead{}&
\colhead{(ksec)}
}
\startdata
BD+60~1704$^D$&9.38&0.372&3.6\nl
BD+59~1757$^W$&9.45&0.259&2.4\nl
BD+60~1703&9.55&0.380&4.9\nl
TYC~4190~2042&11.03&0.383&2.4\nl
TYC~4194~2188&11.15&0.121&5.4\nl
BD+64~1134&11.37&0.044&20.7\nl
TYC~4193~691&11.40&0.182&3.6\nl
TYC~4190~1263$^D$&11.41&0.181&5.4\nl
 
\enddata
\tablecomments{Stars noted as 'D'
  appear as double-lined spectroscopic binaries in the sodium-region spectra, 
  and
  stars noted as 'W' show strong broadening, consistent with a
  spectroscopic binary, but with no indication of duplicity in the
  spectrum. Photometry data are from the Tycho catalog (Hog \etal 1997).}
\end{deluxetable}
 
\vfill
\eject
 
\begin{deluxetable}{ccccc}
\tablewidth{0pc}
\tablecaption{Spectrum Analysis Results}
\tablehead{
\colhead{Star}&
\colhead{Sp.~Class}&
\colhead{Dist.}&
\colhead{EEW}&
\colhead{$v_{abs}$}
}
\startdata
TYC~4193~691&G0V&252$^{+104}_{-74}$&0.23&+4\nl
\vspace{0.1cm}
BD+64~1134&F5V&382$^{+155}_{-110}$&0.05&--12\nl
\vspace{0.1cm}
TYC~4190~1263&F5/G0V&463$^{+192}_{-136}$&0.20&+3 +85\nl
\vspace{0.1cm}
TYC~4194~2188&A6V&618$^{+243}_{-174}$&---&--120\nl
\enddata
\tablecomments{{\it Expected}
  equivalent widths (EEW), in {\AA}ngstroms, of sodium absorption are given {\it relative} to the
  detected absorption in TYC~4194~2188, as detailed in \S 3. Also listed are 
the approximate velocities (in \kmps) of the stellar absorption features in these spectra.}
\end{deluxetable}

\vfill
\eject

\end{document}